%% file: main.tex
\documentclass[sigconf]{acmart}
\usepackage{graphicx}
\usepackage[most]{tcolorbox}
\usepackage{tabularx}
\usepackage{booktabs}
\usepackage{caption}
\usepackage{float}
\usepackage[dvipsnames]{xcolor}

\setcopyright{none}

\acmConference{}{}{}
\acmBooktitle{}
\acmISBN{}
\acmDOI{}
\acmYear{}
\copyrightyear{}

\makeatletter
\renewcommand\@authorsaddresses{}
\makeatother

\begin{document}

\input{title-authors}

\input{abstract/abstract}
\maketitle

\input{introduction}

\input{related-work}

\input{methodology/methodology}

\input{results-analysis/results-analysis}

\input{final-pipeline-analysis/final-pipeline-analysis}

\input{conclusion}

\bibliographystyle{ACM-Reference-Format}
\bibliography{references.bib}

\end{document}

%% file: title-authors.tex
\title{Single-Turn LLM Reformulation Powered Multi-Stage Hybrid Re-Ranking for Tip-of-the-Tongue Known-Item Retrieval}

\author{Debayan Mukhopadhyay}
\affiliation{%
  \institution{Independent Researcher}
  \city{}
  \country{}
}
\email{debayan.mukherjee14@gmail.com}

\author{Utshab Kumar Ghosh}
\affiliation{%
  \institution{Missouri University of Science and Technology}
  \city{}
  \country{}
}
\email{u.ghosh@mst.edu}

\author{Shubham Chatterjee}
\affiliation{%
  \institution{Missouri University of Science and Technology}
  \city{}
  \country{}
}
\email{shubham.chatterjee@mst.edu}
\date{February 2026}

%% file: abstract/abstract.tex
\begin{abstract}
Retrieving known items from vague, partial, or inaccurate descriptions, a phenomenon known as Tip-of-the-Tongue (ToT) retrieval, remains a significant challenge for modern information retrieval systems. 

Our approach integrates a single call to an 8B-parameter Large Language Model (LLM) using a model agnostic prompt for query reformulation and controlled expansion to bridge the gap between ill-formed ToT queries and well-specified information needs in scenarios where Pseudo-Relevance Feedback (PRF) based expansion methods are rendered ineffective, due to poor first stage Recall and Ranking using the raw queries, resulting in expansion using the incorrect documents. 

Importantly, \textbf{the LLM employed in our framework was deliberately kept generic}: \textit{it was not fine-tuned for Tip-of-the-Tongue (ToT) queries, nor adapted to any specific content domains (e.g., movies, books, landmarks). This design choice underscores that the observed gains stem from the formulation of the proposed prompting and expansion strategy itself, rather than from task or domain-specific specialization of the underlying model.} 

Rewritten queries are then processed by a multi-stage retrieval pipeline consisting of an initial sparse retrieval stage (BM25), followed by an ensemble of bi-encoder and late-interaction re-rankers (Contriever, E5-large-v2, and ColBERTv2), cross-encoder re-ranking using monoT5, and a final list-wise re-ranking stage powered by a 72B-parameter LLM (Qwen 2.5 Instruct, 4-bit quantized). 

Experiments on the datasets provided in the 2025 TREC-ToT track show that supplying original user queries directly to this otherwise competitive multi-stage ranking pipeline still yields poor retrieval effectiveness, highlighting the central role of query formulation in ToT scenarios. 

In contrast our light-weight LLM based pre-retrieval query transformation improves the initial Recall net by 20.61\% and then subsequent re-ranking using the re-written queries improves nDCG@10 by 33.88\%, MRR by 29.92\% and Map@10 by 29.98\% over the same pipeline operating on raw queries, indicating that it serves as a highly cost-effective intervention, unlocking substantial performance gains and enabling downstream retrievers and rankers to realize their full potential in Tip-of-the-Tongue retrieval.

All code, data, experimental results and instructions for reproducibility are available at: \textbf{\url{https://github.com/debayan1405/TREC-TOT-2025}}

\end{abstract}

%% file: introduction.tex
\section{Introduction}
Tip-of-the-Tongue (ToT) refers to the process of retrieving information from fragmented, distorted, or incomplete memory. One of the most persistent and difficult problems in cognitive science and information retrieval (IR) is this phenomenon. ToT scenarios are distinguished by a fundamental disconnect between the user's internal representation of an item and the external identifiers needed to locate it, in contrast to traditional known-item search, where a user may recall a title or author imprecisely but retains a clear semantic link to the target \cite{brown1966tot, arguello2021tot}. This disconnect shows up in queries that are frequently verbose, full of phenomenological descriptions of the user's experience (such as "I watched this in my childhood" or "The cover was blue"), and filled with false memories or uncertainty markers \cite{arguello2021tot, lin2023decomposing}.

The field is at a turning point as we get together for the TREC 2025 Tip-of-the-Tongue Track. This track's development, from its original focus on movie identification to the addition of landmarks and celebrities to the more recent open-domain reasoning challenges, reflects the larger evolution of artificial intelligence from pattern matching to agentic reasoning. Early methods relied on dense embeddings to bridge the lexical gap between a user's ambiguous description and a document's metadata, treating ToT retrieval as a specialized case of ad-hoc retrieval \cite{bhargav2022tiptongue}. However, recent scholarship, particularly the introduction of benchmarks such as BLUR (Browsing Lost Unformed Recollections), has demonstrated that semantic similarity alone is insufficient for resolving the most difficult ToT queries \cite{wang2025blur}. Humans solve these problems not simply by matching terms, but by executing complex multi-hop reasoning chains, by validating hypotheses, rejecting false premises and utilizing external tools to bridge gaps in memory \cite{wang2025blur}.

This document summarizes the theoretical foundations and methodological developments that characterize the current state of the art and functions as the foundational Notebook Paper for the 2026 track. We contend that ToT retrieval is no longer just a ranking problem. Rather, it needs to be rethought as an agentic, iterative process of \textit{recollection reconstruction}, in which the retrieval system collaborates with the user to improve their memory trace.

The magnitude of the issue in practical contexts emphasizes how urgent this research is. Analyses of online communities such as Reddit \texttt{/tipofmytongue} reveal millions of unmet information needs and users often resort to human crowdsourcing only when traditional search engines have failed \cite{frobe2023tomtkis, bhargav2022tiptongue}. In both personal and professional contexts, the persistent ToT state is associated with high levels of frustration and cognitive load \cite{arguello2021tot}. In professional settings, the inability to find information again leads to significant productivity losses. Additionally, the democratization of query elicitation through Large Language Models (LLMs) \cite{he2025elicitation} has enabled the creation of large synthetic datasets that can replicate user behavior at previously unachievable scales.

In this introduction, we describe how the ToT problem has persisted in spite of developments in large language models. Although models such as GPT-4 \cite{gpt4} and Claude 3.5 Sonnet show remarkable encyclopedic recall, they often experience hallucinations or are unable to navigate the “uncertainty regions” present in ToT queries \cite{wang2025blur}. In order to assess the \textit{reasoning traces} and \textit{tool-use capabilities} of systems, the 2026 track goes beyond static retrieval effectiveness. Our goal is to create systems that not only retrieve but also \textit{deduce} by combining the most recent architectural advancements in agentic search with insights from cognitive psychology regarding memory access \cite{brown1966tot}.

%% file: related-work.tex
\section{Literature Review}
With concepts from information retrieval, natural language processing, cognitive psychology and library science, the literature on tip-of-the-tongue retrieval is inherently interdisciplinary. The cognitive underpinnings of the phenomenon, the emergence of community-based datasets, the development of specialized retrieval methodologies (from decomposition to simulation), and the current shift toward agentic reasoning benchmarks are the four distinct phases into which this review divides the landscape.

The theoretical basis for the ToT track is rooted in the seminal work of Brown and McNeill \cite{brown1966tot}, who first experimentally induced and described the "Tip of the Tongue" state. They characterized it not as a state of total ignorance, but as a state of \textit{partial access}. Subjects in a ToT state often possessed specific, albeit incomplete, information about the target word, such as its first letter, the number of syllables, or its stress pattern \cite{brown1966tot}. This finding is crucial for IR because it suggests that the "missing" query terms are not randomly absent; rather, users retain structural and sensory fragments of the target while losing the identifying label.

In the context of known-item retrieval, this translates to users recalling "peripheral" attributes rather than "central" identifiers. Arguello et al. expanded this cognitive definition into the digital search domain, conducting a comprehensive annotation study of ToT queries for movies \cite{arguello2021tot}. They identified that searchers rely heavily on three types of information:
\begin{enumerate}
    \item \textbf{Content details:} Descriptions of plot, characters, or scenes.
    \item \textbf{Context of engagement:} Biographical details about when and where the user consumed the content (e.g. I watched this with my father in the 1990s).
    \item \textbf{Previous search attempts:} Meta-commentary on failed strategies (e.g. I already searched for "movie with red car" but found nothing).
\end{enumerate}

Some queries are marked by linguistic hedging. Phrases like "I think", "maybe" or "it might have been" explicitly signal low confidence in specific query facets \cite{he2025elicitation, arguello2021tot}. This uncertainty brings a severe challenge for traditional lexical models like BM25, which treat all query terms with equal weight. Let a user query be "I think the actor was Tom Hanks". In this case, a standard retriever will heavily penalize documents without "Tom Hanks", even if the user's memory is incorrect and the actor was actually Bill Paxton. This phenomenon of "false memories" is a pervasive feature of ToT queries and requires systems that can treat query terms as probabilistic constraints rather than absolute requirements \cite{hauff2012realistic}. The linguistic structure of these queries further complicates retrieval. 

Unlike the terse keywords typical of navigational search, ToT queries are notoriously verbose. Analysis of the \textit{WhatsThatBook} dataset reveals an average query length of 156.20 words, vastly exceeding the 7-20 word averages found in standard retrieval benchmarks like MS MARCO or Natural Questions \cite{lin2023decomposing}. This verbosity dilutes the signal of key terms, as users often bury critical details within lengthy narratives about their search process or emotional state. 

\subsection{The Evolution of Datasets}

Progress in ToT retrieval has been historically bottlenecked by data scarcity. Unlike ad-hoc retrieval, where relevance judgments can be crowd-sourced, ToT queries require a ground truth that only the original searcher can verify. The evolution of datasets reflects a migration from small, hand-curated collections to massive, community-mined and synthetically generated corpora.

\subsubsection{\textbf{Community Question Answering (CQA) :}}
 Mining Community Question Answering (CQA) platforms, particularly the \texttt{r/tipofmytongue} subreddit, led to the first breakthrough in data availability. 
 \begin{itemize} 
    \item \textbf{Reddit-TOMT:}  This method was formalized by Bhargav et al., who created the Reddit-TOMT dataset \cite{bhargav2022tiptongue} by scraping solved threads from Reddit. 15,000 query-item pairs with a movie and book focus were included in this dataset. It determined the task's baseline difficulty and demonstrated that neural retrievers such as DPR (Dense Passage Retrieval) greatly outperformed lexical baselines because of the semantic difference between the formal metadata of the items and the user's verbose description \cite{bhargav2022tiptongue}.

    \item \textbf{TOMT-KIS:}  This method was greatly expanded by Frobe et al., who published the TOMT-KIS dataset with 1.28 million known-item questions \cite{frobe2023tomtkis}.  Only 47\% of the queries had identified answers and the emphasis was primarily on Query Performance Prediction (QPP) rather than pure retrieval effectiveness, despite the dataset's unmatched scale.  By using the term "time to solve" as a stand-in for query difficulty, their analysis showed that known-item questions have different performance characteristics than typical informational queries \cite{frobe2023tomtkis}.
    
     \item \textbf {WhatsThatBook:}  Further specializing, Lin et al. presented the WhatsThatBook dataset, which consists of 14,441 pairs created especially for book retrieval \cite{lin2023decomposing}.  This dataset was special because it included rich metadata fields, such as images of book covers, which made multimodal retrieval techniques possible. It highlighted that users often recall visual features ("a yellow cover with a cat silhouette") that textual descriptions in standard indices fail to capture \cite{lin2023decomposing}.
\end{itemize}

\subsubsection{\textbf{The Synthetic Turn: Elicitation and Simulation :}}
Despite the volume of CQA data, these datasets suffer from inherent domain skew. As noted by He et al., CQA platforms are heavily biased toward casual leisure topics such as movies, games and music, leaving domains like "People" or "Landmarks" underrepresented \cite{he2025elicitation}. To address this, the field has moved toward \textit{query elicitation} and \textit{simulation}.

He et al. introduced a dual-method framework for generating ToT queries \cite{he2025elicitation}. First, they employed \textbf{LLM-based Simulation}. By prompting Large Language Models (LLMs) with Wikipedia summaries and specific instructions to "role-play" a forgetful user (e.g. "express doubt", "mix up details"), they generated synthetic queries that achieved high correlation with real human queries in terms of system ranking performance. Second, they utilized \textbf{Human Elicitation via Visual Stimuli}. To capture genuine human memory failures, they developed an interface that showed participants images of famous entities (movies, landmarks) and asked them to recall the name. If the participant recognized the entity but failed to name it, they were in a genuine ToT state and were asked to write a query.

\subsection{Methodological Advances: Decomposition and Multimodality}

The complexity of ToT queries renders standard single-vector retrieval ineffective. A single embedding vector struggles to encode the "plot", "cover description", "time period" and "user sentiment" simultaneously without losing fidelity. This has led to the adoption of \textit{Query Decomposition} and \textit{Facet-Specific Retrieval}.

\subsubsection{\textbf{Query Decomposition :}}
Lin et al. proposed a framework that decomposes complex ToT queries into individual "clues" or sub-queries, which are then routed to specialized retrieval experts \cite{lin2023decomposing}. For instance, a query mentioning "a yellow cover" and "a plot about a boy and a cat" is split. A "Cover Retriever" (using CLIP) processes the visual description, while a "Plot Retriever" (using Contriever) processes the narrative elements.

They distinguished between \textbf{Extractive} and \textbf{Predictive} decomposition. Their results showed that \textbf{Predictive Decomposition} significantly outperformed extractive methods, as it bridges the gap between the user's \textit{experience} (context) and the document's \textit{metadata} (facts) \cite{lin2023decomposing}.

\subsubsection{\textbf{Multimodal Integration :}}
The integration of visual data has become increasingly central. The \textit{WhatsThatBook} dataset demonstrated that cover image retrieval (via CLIP) could resolve queries that were textually ambiguous \cite{lin2023decomposing}. Similarly, the BLUR benchmark contains file uploads (images, audio clips) as query inputs, which shows the reality that users often possess a visual fragment (e.g. a screenshot of a scene) but lack the semantic label \cite{wang2025blur}.

\subsection{The Agentic Shift: Reasoning and Benchmark Validity}

The most significant recent development, driving the agenda for TREC 2026, is the realization that ToT retrieval is often a \textbf{multi-hop reasoning problem} rather than a static matching problem. Wang et al. introduced the \textbf{BLUR (Browsing Lost Unformed Recollections)} benchmark to test this capability rigorously \cite{wang2025blur}.

\subsubsection{\textbf{Limitations of Static Retrieval :}} BLUR showed that even the most advanced static retrievers and base LLMs (such as GPT-4o \cite{gpt4o} or Llama 3.1) fall well short of near-perfect human performance (98\%) on complex ToT questions, achieving only $\sim$56\% accuracy \cite{wang2025blur}.  Models have trouble with \textbf{orchestration} and \textbf{contextual verification}, according to the failure case analysis.

\subsubsection{\textbf{Tool Use and Agentic Systems :}}
 A change toward \textbf{Agentic Search} has been required as a result. According to this paradigm, the system is an active agent capable of using tools (browsers, maps, OCR, calculators) rather than a passive retriever.  The BLUR benchmark specifically assesses how well a system can plan a search strategy, carry out actions and compare the results to the user's ambiguous description \cite{wang2025blur}.
 However, Wang et al. discovered an unexpected "Agentic Gap" - complex agent frameworks frequently only slightly outperformed the base LLMs they wrapped \cite{wang2025blur}.  This implies that the fundamental challenge is not only having access to tools but also having the \textbf{reasoning capability} needed to deal with uncertainty.

%% file: methodology/methodology.tex
\section{Methodology}
\input{methodology/motivation}
\input{methodology/multi-model-query-rewriting}
\input{methodology/sparse-retrieval}
\input{methodology/dense-retrieval}
\input{methodology/cross-encoder}
\input{methodology/llm-listwise-reranking}

%% file: methodology/motivation.tex
\subsection{Motivation: The Challenge of ToT Artifacts}
Our initial analysis of the queries provided by the TREC-TOT 2025 Track Organizers \cite{trectot2025guidelines} revealed significant challenges for current State-of-the-Art (SoTA) Retrieval-Augmented Generation (RAG) systems. The original queries often contain what we term "ToT artifacts", which are verbose, circumlocutory descriptions characteristic of the Tip-of-the-Tongue phenomenon, where the user describes an item without naming it directly. These artifacts cause a vocabulary mismatch between the queries and the target document in the corpus, which is composed of keyword rich metadata.

We established this finding quantitatively by analyzing the baseline performance. As shown in (Table~\ref{tab:baseline-performance}), the original queries yield suboptimal retrieval performance across all dataset splits. This motivated our adoption of a query rewriting mechanism to transform these artifact-laden descriptions into more precise, keyword-centric queries suitable for efficient retrieval.

\textbf{Our methodology is guided by the premise} that tip-of-the-tongue (TOT) retrieval is fundamentally a \textbf{problem of cognitive signal reconstruction rather} than conventional document ranking. TOT queries typically encode partial, uncertain, and highly verbose memory traces that are poorly aligned with the assumptions of standard retrieval systems. Accordingly, we adopt a query-centric design philosophy, treating query interpretation and reformulation as first-class components of the retrieval pipeline. 

Specifically, we introduce a structured, LLM-driven reformulation process that sequentially transforms raw TOT queries into semantically grounded representations suitable for downstream retrieval. This process includes controlled entity identification and expansion embedded directly within the prompt, with explicit guardrails to preserve faithfulness to the original user intent and mitigate hallucination. Rather than injecting externally retrieved entities or aligning to corpus-specific identifiers, expansion is performed at the level of query cognition, allowing the system to remain agnostic to the structure and contents of any particular collection.

Importantly, this design choice also avoids reliance on external knowledge bases for entity expansion. Conventional entity expansion typically involves linking queries to canonical identifiers in resources such as Wikidata \cite{10.1145/2629489} or Google Knowledge Graph \cite{google_knowledge_graph_api}, which implicitly couples retrieval performance to the coverage and assumptions of these public information domains. Such coupling is incompatible with realistic ToT scenarios, where targets may be private, proprietary, emerging, or otherwise absent from curated knowledge graphs. In these settings, controlled expansion based strictly on information present in the user query is the only viable alternative. Accordingly, our approach performs guardrailed entity and vocabulary expansion entirely within the prompt, restricting additions to aliases, canonical forms, and domain descriptors that are directly supported by the input, and abstaining from expansion when ambiguity remains.

This design contrasts with corpus-centric retrieval paradigms, which emphasize adapting models to a fixed dataset through domain-specific indexing, finetuning, or synthetic supervision. While such approaches can be highly effective when the target domain and corpus are known a priori, they inherently couple system performance to the stability and characteristics of that collection. In contrast, our approach deliberately avoids corpus-dependent training signals and assumes an open-world setting in which domains are unknown and document collections may evolve dynamically (e.g., web-scale retrieval). 

All large language models in our pipeline are used entirely off-the-shelf and in a zero-shot manner, without any task-specific finetuning or supervised adaptation. System behavior is instead governed through carefully designed prompting, enabling controlled query reconstruction without reliance on closed feedback loops or domain-tailored supervision.

Conceptually, we frame TOT retrieval as a two-stage process: (i) cognitive reconstruction of the latent target intent from noisy natural-language memory cues, followed by (ii) retrieval and ranking over candidate documents. By prioritizing the former, our method emphasizes robustness under distribution shift and transferability across arbitrary domains.

This framing allows the same system to operate over heterogeneous corpora without modification, treating the LLM primarily as a reasoning and interpretation component rather than as a domain-adapted retriever. As a result, our pipeline is designed to generalize beyond benchmark-specific conditions, supporting realistic deployment scenarios where corpus characteristics are unknown or continuously changing, and where retraining or finetuning on the target collection is infeasible.
\input{tables/baseline-performance}

%% file: tables/baseline-performance.tex
\begin{table}[t]
\centering
\caption{Baseline retrieval performance across dataset splits using the original queries}
\label{tab:baseline-performance}
\small
\begin{tabular}{c c c c}
\hline
\textbf{Dataset Split} & \textbf{nDCG@10} & \textbf{Recall@1000} & \textbf{MRR} \\
\hline
Train & 0.0585 & 0.4475 & 0.0537 \\
Dev1  & 0.0834 & 0.4437 & 0.0806 \\
Dev2  & 0.0840 & 0.4476 & 0.0696 \\
Dev3  & 0.3366 & 0.7705 & 0.3129 \\
Test  & 0.1223 & 0.4855 & 0.1114 \\
\hline
\end{tabular}
\end{table}

%% file: methodology/multi-model-query-rewriting.tex
\subsection{Multi-Model Query Rewriting}
To mitigate the influence of ToT artifacts, we adopt a diverse rewriting strategy using three instruction-tuned large language models: \textbf{Meta-Llama-3.1-8B-Instruct}~\cite{llama}, \textbf{Mistral-7B-Instruct-v0.3}~\cite{mistral}, and \textbf{Qwen2.5-7B-Instruct}~\cite{qwen}. All models are deployed at their maximum supported BF16 precision and decoded using greedy decoding ($\texttt{do\_sample}=\mathrm{False}$), with temperature set to $0.0$, nucleus sampling threshold $\texttt{top\_p}=0.95$ (reported for completeness), and $\texttt{max\_new\_tokens}=128$. All decoding parameters are fixed across models to ensure deterministic generation and facilitate reproducibility.

The core of our strategy involves directing these LLMs to strip away the conversational fluff and focus on the semantic core of the user's intent. We utilized the following specific prompt to guide the rewriting process which forces the models to act as a strict filter, distilling the user's vague recollection into a structured search query.:

\begin{tcolorbox}[
  title={Rewriter Prompt},
  colback=gray!5,
  colframe=black!60,
  fonttitle=\bfseries,
  breakable,
  width=\columnwidth,
  listing only,
  listing options={
    basicstyle=\ttfamily\footnotesize,
    breaklines=true,
    breakatwhitespace=true,
    columns=fullflexible,
    keepspaces=true
  }
]
\textbf{System Prompt:} You are a query rewriter for a search engine. Your task is to rewrite a complex, verbose, tip-of-the-tongue description into a simple, keyword-focused search query.
\\\\
\textbf{Guidelines:}
1. Identify the core entity type if implied (e.g., could be a movie, book, song, product, place, person, software tool, concept, etc. - the domain is open-ended). Perform cautious entity expansion by adding closely related aliases or canonical forms only if the anchor entity is directly mentioned or supported by the input text. Never guess, invent, or infer entities that are not explicitly mentioned or unambiguously implied. If unsure, do not expand.\\

2. Incorporating domain specific vocabulary (e.g., movies: ``cinematography'', ``anthology film''; books: ``epistolary novel'', ``bildungsroman''; science: ``biochemical pathway'', ``quantum phenomenon''; products: ``form factor'', ``backwards compatibility''). 

3. Extract key details in a domain-agnostic way: concrete attributes such as events, functions, features, relationships, names, dates, locations, behaviors, or other unique identifiers explicitly stated in the input (e.g., plot points for media, specifications for products, symptoms for medical queries, APIs for software, etc.). Do not add new facts.\\

4. Remove conversational filler ("I think it was...", "It might be...", "I remember seeing...").\\

5. Remove negative constraints or uncertainty unless crucial ("Not sure if...").\\

6. \textbf{Strict grounding rule:} Do NOT introduce any information, entities, attributes, or assumptions that are not present in the input query. Every token in the rewritten query must be traceable to the original text or to safe lexical transformations (e.g., synonyms). No external knowledge.\\

7. Formulate a concise query that a standard search engine (like Google or BM25) would understand. Output ONLY the rewritten query text. Do not output any explanations.\\
\end{tcolorbox}

%% file: methodology/sparse-retrieval.tex
\subsection{First-Stage Sparse Retrieval (Stage 1)}\label{sec:sparse_parameter_tuning}
To establish a strong sparse retrieval baseline, we employed the Okapi BM25 \cite{10.1561/1500000019} and implemented a rigorous, sequential hyperparameter optimization strategy to maximize retrieval performance, specifically targeting Recall@1000.

\subsubsection{\textbf{Sequential Adaptive Grid Search :}}
\label{sec:adaptive-grid-search-tuning}
We adopted a multi-stage tuning approach that progressed sequentially through our dataset splits (\texttt{train} $\rightarrow$ \texttt{dev1} $\rightarrow$ \texttt{dev2}).

This strategy allows for the refinement of parameters on unseen data while carrying forward the "knowledge" (optimal parameters) from the previous stage, thereby preventing overfitting to a single development set.

The optimization process was divided into two distinct phases:

\textbf{1. Phase I: Global Search (Training Set) :}
On the initial training set, we conducted a coarse-grained grid search over a broad range of values to identify the global optimal region. The search spaces were defined as follows:
\begin{itemize}
    \item \textbf{BM25:} We explored $k_1 \in [0.4, 4.0]$ and $b \in [0.3, 1.0]$
\end{itemize}

\textbf{2. Phase II: Refined Local Search (Development Sets) :}
For subsequent datasets (\texttt{dev1}, \texttt{dev2}), we employed a refined local search strategy. Instead of searching the global grid again, we generated a narrower, high-resolution grid centered around the best parameters found in the previous stage.
Let $\theta_{best}$ be the optimal value for a parameter from the previous dataset. The new search space $S$ was constructed as:
\begin{equation}
    S = \text{linspace}(\theta_{best} - \delta, \theta_{best} + \delta, \text{steps}=5)
\end{equation}
where $\delta$ represents the refinement radius ($\delta_{k1}=0.4$, $\delta_{b}=0.2$).

The optimization pipeline was implemented using the PyTerrier framework. To handle the computational load of evaluating hundreds of configurations over the 6.4M document corpus, we utilized a high-performance computing environment with 500GB of RAM. We employed parallel processing via \texttt{ThreadPoolExecutor}, allowing for concurrent evaluation of multiple parameter configurations. The index structures were loaded into memory using the \texttt{fileinmem} property to minimize I/O latency during the iterative retrieval process.

%% file: methodology/dense-retrieval.tex
\subsection{Bi-encoder and Late Interaction Based Re-ranking (Stage 2)}
Following the initial sparse retrieval, we employed traditional bi-encoders and ColBERTv2's Late Interaction mechanism to address the vocabulary mismatch inherent in Tip-of-the-Tongue queries. This stage was designed as a high-recall filter, aggregating semantic signals from multiple architectures. We would also like to point out that, even for re-ranking, the queries rewritten by the Mistral model, is being used.

\subsubsection{\textbf{Ensemble Architecture :}}
\begin{enumerate}
    \item \textbf{ColBERTv2.0:} A late-interaction model \cite{santhanam-etal-2022-colbertv2} that retains token-level granularity, effective for capturing fine-grained details in verbose ToT queries.
    \item \textbf{Contriever:} An unsupervised dense retriever trained via contrastive learning \cite{izacard2021contriever}, offering robust zero-shot performance.
    \item \textbf{E5-Large-v2:} A text-embedding model \cite{wang2024textembeddingsweaklysupervisedcontrastive} trained with weak supervision, providing strong semantic matching capabilities.
\end{enumerate}

The outputs of these three models were combined using Reciprocal Rank Fusion (RRF) with a RRF constant $k=60$ to produce a single, robust candidate list.

\subsubsection{\textbf{Robust Candidate Depth Tuning :}}
\label{robust-candidate-depth-tuning}
A critical design choice was the selection of the candidate depth $K_{dense}$. Unlike standard approaches that optimize for nDCG, we optimized for \textbf{Recall@1000} again to ensure the relevant document was retained for the downstream stages. To avoid overfitting to the training set, we employed a \textit{Robust Aggregate Tuning} strategy. We defined a robust score $S_K$ for each candidate depth $K$:
\begin{equation}
    S_K = \mu(R@K) - \alpha \cdot \sigma(R@K)
\end{equation}
where $\mu$ and $\sigma$ are the mean and standard deviation of Recall@K across the \texttt{train}, \texttt{dev1}, and \texttt{dev2} splits, and $\alpha=0.5$ is a stability penalty.

To implement this efficiently, we utilized a "Virtual Tuning" approach: we ran inference once to generate a master list of the top-5000 candidates and subsequently simulated smaller $K$ values by slicing this list.

%% file: methodology/cross-encoder.tex
\subsection{Cross-Encoder Re-ranking (Stage 3)}
The Bi-encoder and Late Interaction Based Re-ranking Stage, while improving recall, introduced precision errors by prioritizing broad semantic matches over specific keyword constraints. To correct this "semantic drift", we implemented a Cross-Encoder stage using the \texttt{monoT5-large} \cite{nogueira2020documentrankingpretrainedsequencetosequence} model.

\subsubsection{\textbf{Hybrid Input Pooling :}}
To maximize the probability of including the relevant document, we devised a hybrid input strategy. Rather than re-ranking only the dense output, we constructed a union pool consisting of the top-100 candidates from the sparse BM25 stage and the top-100 candidates from the RRF performed in the previous stage. This ensured that the Cross-Encoder had access to both exact keyword matches and semantic matches.

\subsubsection{\textbf{Implementation and Optimization :}}
The \texttt{monoT5} model processes query-document pairs jointly, which is computationally intensive. We optimized throughput by enabling FP16 (half-precision) inference and distributing the workload across two NVIDIA RTX 6000 GPUs using \texttt{DataParallel}.

%% file: methodology/llm-listwise-reranking.tex
\subsection{Large Language Model Re-ranking (Stage 4)}
The final stage employed a $72$B-parameter large language model,
$\texttt{Qwen2.5-72B-Instruct-AWQ}$,
served via vLLM~\cite{qwen,qwen2.5}, to perform zero-shot listwise re-ranking. This component acted as the final arbiter, leveraging the model’s world knowledge to resolve subtle ambiguities in ToT descriptions.

\paragraph{\textbf{Decoding details :}}
We used greedy decoding with sampling disabled (i.e., $\texttt{do\_sample} = \texttt{False}$), yielding fully deterministic outputs for identical inputs. The maximum generation length was set to $\texttt{max\_new\_tokens} = 128$. Since greedy decoding was employed, $\texttt{temperature}$, $\texttt{top\_p}$, and $\texttt{top\_k}$ were not applied. The repetition penalty was set to its default value of $\texttt{repetition\_penalty} = 1.0$, with no additional presence or frequency penalties. The padding token was aligned with the end-of-sequence token (i.e., $\texttt{pad\_token\_id} = \texttt{eos\_token\_id}$). Under this configuration, the model selects the highest-probability token at each decoding step.

\subsubsection{\textbf{Listwise Ranking Strategy :}}

We employed a list-wise approach, in which the LLM is presented with the query and a batch of candidates simultaneously and prompted to output the optimal permutation of Document IDs. This method allows the model to compare candidates directly against each other, rather than scoring them in isolation. 
\\
A manual inspection of Wikipedia pages revealed that, the Introduction section of these web pages are usually in the range 450-600 characters. To fit within the context window while retaining critical information, document text was truncated to the first 500 characters, capturing the introductory summaries typical of Wikipedia articles.
\\
While it is highly likely that the queries might ask for information buried deeper inside the full text, supplying the full text as the context window has the following issues:
\\
\begin{itemize}
\item \textbf{Cost/Speed:} Processing full documents (e.g., 2000 tokens) for 50 candidates would explode the input size to 100k+ tokens per query. This would make the 72B model extremely slow, essentially moving from seconds to minutes to process each query, and could also confuse it with irrelevant details.
\\
\item \textbf{"Lost in the Middle":} LLMs often focus on the beginning and end of long contexts. Truncating ensures the model focuses on the most "dense" information (the start). ToT queries are often descriptive and conceptual (e.g., "movie where guy is stuck on Mars growing potatoes"). The relevant document (e.g., a Wikipedia plot summary) usually contains the key matching concepts early in the text (title, introduction, first paragraph).
\end{itemize}

\begin{tcolorbox}[
  title={List-wise Re-ranker Prompt},
  colback=gray!5,
  colframe=black!60,
  fonttitle=\bfseries,
  breakable,
  width=\columnwidth,
  listing only,
  listing options={
    basicstyle=\ttfamily\footnotesize,
    breaklines=true,
    breakatwhitespace=true,
    columns=fullflexible,
    keepspaces=true
  }
]
You are an expert search relevance ranker.
Your task is to re-rank the following candidate documents based on their relevance to the user query.
The goal is to place the true relevant document at the very top (Rank 1).

Query: {query}

Candidates:
{candidates}

\textbf{Instructions:}\\
1. Analyze the query and the candidates carefully.\\
2. Output the ranking as a list of IDs in order of relevance, from most relevant to least relevant.\\
3. Use the format: [ID] > [ID] > [ID] ...\\
4. Only output the ranking, no explanation.\\

Ranking:
\end{tcolorbox}

\subsubsection{\textbf{Joint Hyperparameter Tuning :}}
Recognizing the dependency between the Cross-Encoder and the LLM, we performed a joint grid search to optimize the input depth ($K_{cross}$) fed to the LLM and the final re-ranking depth ($K_{llm}$).

\begin{figure}[h]
\centering
\fbox{\includegraphics[width=0.4\textwidth]{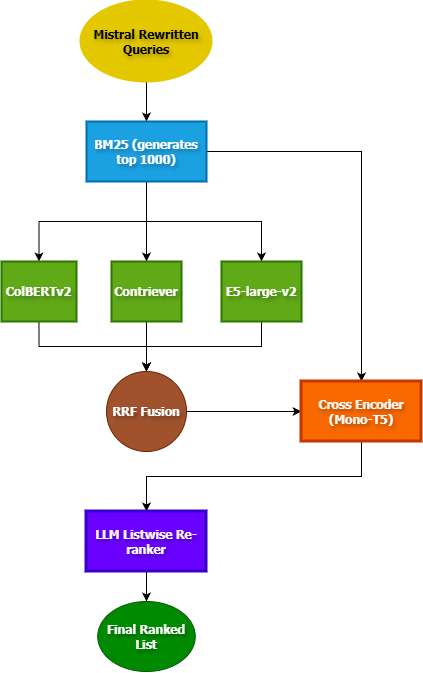}}
\caption{The proposed Retrieval Pipeline with four distinct stages: the Sparse Retrieval Stage (Stage 1), Dense Retrieval Bi-encoder (Stage 2), Cross-Encoder Stage (Stage 3), and the LLM Re-ranker Stage (Stage 4). 
The input re-written query set in the diagram has been generated using Mistral-7B-Instruct-v0.3, as discussed in Section~\ref{sec:results_analysis}.}
\label{fig:retrieval_pipeline}
\end{figure}

%% file: results-analysis/results-analysis.tex
\section{Results and Analysis}\label{sec:results_analysis}

\input{results-analysis/llm-rewriting-efficacy}

\input{results-analysis/bi-encoder-analysis}

\input{results-analysis/cross-encoder-analysis}

\input{results-analysis/llm-listwise-reranking-analysis}

%% file: results-analysis/llm-rewriting-efficacy.tex
\subsection{Efficacy of LLM-based Query Rewriting}

The introduction of an LLM-based rewriting step at the beginning of the retrieval pipeline introduces a non-trivial computational cost and latency overhead. Therefore, the primary question we must address is whether this additional complexity translates into justifiable performance gains.

To evaluate this, we create an ablation environment, where we fix the first stage retriever to BM25 and analyze the rewrites of three distinct Large Language Models (LLMs)-Llama, Mistral, Qwen against the baseline of the original Tip-of-the-Tongue (ToT) queries.

It is important to establish our evaluation protocol regarding the dataset splits. The \texttt{train}, \texttt{dev1}, and \texttt{dev2} splits were utilized exclusively to fine-tune the hyperparameters of the retrieval models (specifically $k1$ and $b$ for BM25), a process detailed in \textbf{(Section \ref{sec:sparse_parameter_tuning})}. Consequently, the \texttt{dev3} split functions as a pseudo-test set, representing unseen data where no parameter optimization was performed. This distinction ensures that the results presented on \texttt{dev3} reflect generalized performance.

\subsubsection{\textbf{Failure of Pseudo Relevance Feedback (PRF) :}}

\paragraph{\textbf{Pseudo-Relevance Feedback (PRF) settings:}}
Unless otherwise specified, all query expansion methods use PyTerrier defaults: Bo1, KL, and RM3 are configured with 3 feedback documents and 10 expansion terms. RM3 additionally uses $\lambda=0.6$, controlling the weight of the relevance model in query interpolation.

\paragraph{\textbf{Discussion:}}
Contrary to expectations, all three pseudo-relevance feedback methods (RM3, Bo1, and KL divergence) underperform the BM25 baseline across every evaluation metric \textbf{(Table \ref{tab:rewrite_efficacy})}. Most notably, their \textbf{Recall@1000} values collapse substantially (RM3: 0.5093, Bo1: 0.7295, KL: 0.7332) compared to the original queries (0.7705), indicating a severe degradation in first-stage retrieval coverage.

This behavior can be attributed to the inherent difficulty of Tip-of-the-Tongue (ToT) queries, which are typically sparse, noisy, and semantically incomplete. Under such conditions, the initial BM25 rankings are often unreliable, causing PRF methods to expand queries using predominantly non-relevant documents. Rather than correcting the query intent, this process amplifies retrieval drift, leading to degraded recall and ranking quality.

This failure mode is further exacerbated by the shallow relevance judgments of the TREC ToT 2025 collection, which effectively represents a single-relevant-document setting. In such a regime, even minor errors in early-stage retrieval are catastrophic: once the sole relevant document is missed in the top-$k$ candidates, PRF mechanisms inevitably reinforce incorrect topical signals.

As a result, PRF-based expansion not only fails to recover missing relevance but actively suppresses it, producing recall losses of up to 26\% relative to the baseline (RM3 vs.\ Original). These findings highlight a fundamental limitation of traditional feedback approaches in low-recall, single-relevance scenarios and motivate the need for semantically informed reformulation mechanisms.

\subsubsection{\textbf{Performance Comparison and Model Selection :}}

The results on the pseudo-test set (\texttt{dev3}) clearly demonstrate that Query Rewriting provides a massive performance boost compared to the original queries.

As illustrated in \textbf{(Table \ref{tab:rewrite_efficacy})}, beyond improvements in ranking-oriented metrics such as nDCG@10 and MRR, LLM-based rewriting also delivers consistent gains in \textbf{Recall@1000}, which is critical for downstream re-ranking pipelines. While the original queries achieve a Recall@1000 of 0.7705, Mistral increases this to 0.8284, representing a substantial expansion of the candidate set containing the relevant document. This enhanced first-stage recall directly enables more effective second-stage re-ranking, explaining the large improvements observed in Success@100 and MAP@10.

Importantly, this behavior contrasts sharply with PRF-based approaches, whose reduced recall severely limits any potential gains from subsequent ranking models. Among the evaluated LLMs, Mistral demonstrates the strongest ability to recover latent query intent, producing both the highest recall and the best overall retrieval effectiveness. Consequently, we adopt Mistral as the default query rewriter for all subsequent experiments.

\subsubsection{\textbf{Rewrite Reproducibility Analysis}}

\paragraph{\textbf{Motivation.}}
LLM-based query reformulation can introduce stochasticity, raising concerns about whether observed improvements reflect consistent model behavior or isolated favorable generations. To assess reproducibility, we conduct an \emph{Rewrite Reproducibility Analysis} using Recall@1000 as the target metric, quantifying variability across repeated reformulation runs.

\paragraph{\textbf{Experimental Setup.}}
We deterministically sampled 100 queries from the test set using a fixed random seed (SEED=42). For each query, we performed three independent reformulation runs with identical decoding parameters. All experiments were conducted using the Mistral-7B-Instruct-v0.3 model at its maximum supported BF16 precision.

Generation was performed using greedy decoding ($\texttt{do\_sample}=\mathrm{False}$), with temperature set to $0.0$, nucleus sampling threshold $\texttt{top\_p}=0.95$ (reported for completeness), and $\texttt{max\_new\_tokens}=128$. Each reformulated query was evaluated independently, and $\mathrm{Recall@1000}$ was computed per run before aggregating statistics across all sampled queries.

Because decoding is fully deterministic under this configuration, we do not report results across multiple random seeds: varying the seed does not affect generation outcomes in the absence of sampling. All decoding parameters are fixed across runs to ensure deterministic behavior and facilitate reproducibility.

\paragraph{\textbf{Metrics:}}
We report mean Recall@1000 ($\mu$), standard deviation ($\sigma$), and coefficient of variation (CV), defined as:

\begin{equation}
\mathrm{CV} = \frac{\sigma}{\mu} \times 100\%.
\end{equation}

The CV provides a normalized measure of dispersion relative to overall performance.

\paragraph{\textbf{Results:}}
The Recall@1000 values across runs were \{0.71, 0.70, 0.75\}, yielding a mean of 0.7200 and a standard deviation of 0.0265, corresponding to a CV of 3.67\%. Table~\ref{tab:output_sensitivity} summarizes these results.

\begin{table}[t]
\centering
\caption{Rewrite Reproducibility analysis of LLM-based query reformulation across three runs on 100 sampled test queries.}
\label{tab:output_sensitivity}
\begin{tabular}{lcccc}
\hline
\textbf{Model} & \textbf{Runs} & \textbf{Mean R@1000} & \textbf{Std. Dev.} & \textbf{CV (\%)} \\
\hline
Mistral & 3 & 0.7200 & 0.0265 & 3.67 \\
\hline
\end{tabular}
\end{table}

\paragraph{\textbf{Discussion:}}
Based on our predefined criteria, CV values below 2\% indicate high stability, while values between 5--10\% require explicit reporting of variability. The observed CV of 3.67\% falls within the moderate range, indicating limited sensitivity to stochastic generation effects. We therefore report mean $\pm$ standard deviation and conclude that performance gains are reproducible across runs rather than attributable to isolated LLM outputs. Residual variability is acknowledged as a limitation.

\subsubsection{\textbf{Query Type Analysis}}

\paragraph{\textbf{Motivation:}}
We analyze performance by query type to understand where our proposed reformulation method excels and where it fails versus the original queries which we treat as the baseline.

\paragraph{\textbf{Methodology:}}
We deterministically sampled (SEED = 42) and manually labeled 100 queries (from 622 total test queries) according to difficulty and intent. The resulting categories were:

\begin{itemize}
    \item \textbf{Simple entity recall}: e.g., ``A book by Malcolm Gladwell about 10{,}000 hours''
    \item \textbf{Complex constraints}: e.g., ``Sci-fi movie, 90s, time loop, NOT Groundhog Day''
    \item \textbf{False memory}: e.g., ``That song by Nirvana with the violin solo''
    \item \textbf{Visual or descriptive recall}: e.g., ``Painting with melting clocks''
\end{itemize}

For each category, we compute \textbf{Recall@1000 (R@1000)}

\paragraph{\textbf{Results:}}

Our proposed method shows consistent gains on higher-level semantic queries. The largest improvement appears on \textit{False Memory} queries (+0.29 R@1000), suggesting stronger robustness to incorrect or misleading premises. Substantial gains are also observed for \textit{Visual or Descriptive Recall} (+0.15) and \textit{Complex Constraints} (+0.09), indicating improved compositional and descriptive reformulation.

In contrast, performance degrades on \textit{Simple Entity Recall} (--0.20), where the baseline already performs strongly. This suggests that the method primarily benefits ambiguous, compositional, or concept-heavy queries, while offering limited advantage for straightforward factual retrieval.

\begin{table*}[h]
\centering
\small
\caption{Recall@1000 by query type. Underlying retrieval was performed using BM25}
\label{tab:query-type-analysis}
\begin{tabular}{lcccc}
\hline
\textbf{Category} & \textbf{Count} & \textbf{Baseline R@1000} & \textbf{Mistral R@1000} & \textbf{$\Delta$ R@1000} \\
\hline
Complex Constraints & 34 & 0.5294 & 0.6176 & +0.0882 \\
False Memory & 7 & 0.4286 & 0.7143 & +0.2857 \\
Simple Entity Recall & 5 & 0.8000 & 0.6000 & --0.2000 \\
Visual / Descriptive Recall & 54 & 0.3704 & 0.5185 & +0.1481 \\
\hline
\end{tabular}
\end{table*}

\paragraph{\textbf{Discussion:}}

Overall, these results indicate that improvements are concentrated in harder query classes requiring abstraction or constraint satisfaction, rather than direct entity lookup.

\subsubsection{\textbf{Domain Centric Vocabulary Injection Analysis}}

\paragraph{\textbf{Motivation:}}
Effective query reformulation should ideally introduce domain-appropriate terminology, as specialized vocabulary can improve retrieval by better aligning queries with corpus-specific language. We therefore analyze whether the LLM produces domain-specific phrasing or defaults to generic paraphrasing during reformulation.

\paragraph{\textbf{Methodology:}}
Using the same deterministically sampled test subset (SEED = 42), we manually inspected 100 reformulated queries and categorized each output according to vocabulary usage:

\begin{itemize}
    \item \textbf{Domain-Specific}: reformulations include specialized terminology appropriate to the query domain (e.g., movies: ``cinematography'', ``anthology film''; books: ``epistolary novel'', ``bildungsroman''; science: ``biochemical pathway'', ``quantum phenomenon''; products: ``form factor'', ``backwards compatibility'');
    \item \textbf{Generic}: reformulations rely on nonspecific constructions (e.g., ``a movie about'', ``a book where'').
\end{itemize}

We report counts and fractions per category. Let $Y$ denote the number of generically paraphrased queries; we report the generic rate as $Y/100$.

\paragraph{\textbf{Results:}}
As shown in Table~\ref{tab:domain_vocab}, 63\% of the reformulations employed domain-specific vocabulary, while 37\% relied on generic paraphrasing.

\begin{table}[t]
\centering
\caption{Domain vocabulary usage in reformulated queries on the deterministically sampled test subset (SEED = 42).}
\label{tab:domain_vocab}
\begin{tabular}{lcc}
\hline
\textbf{Category} & \textbf{Fraction} \\
\hline
Domain-Specific & 63\% \\
Generic & 37\% \\
\hline
\end{tabular}
\end{table}

\paragraph{\textbf{Discussion:}}
The model introduces domain-specific terminology in a majority of cases, suggesting that the reformulation strategy often enriches queries with specialized language rather than relying solely on generic templates. However, over one-third of outputs remain generic, indicating room for improvement in consistently leveraging domain knowledge.

\subsubsection{\textbf{Guard-railed Entity Expansion in Resolving Indirect References}}

\paragraph{\textbf{Motivation:}}
To further analyze the behavior of LLM-driven query reformulation under indirect references, we conduct an auxiliary study on indirect reference resolution. Indirect references arise when a query refers to a target entity only through a relational description, for example, “the author of [book]”, “the actor from [movie]”, or “the company that makes [product]”, rather than naming the entity explicitly. Such constructions are common in ToT queries and pose a critical challenge for entity expansion mechanisms.

\paragraph{\textbf{Methodology:}}
We evaluate reformulation outputs along two dimensions. First, we identify whether a query contains an indirect entity pointer. Second, we categorize the reformulation behavior into one of four outcomes: Resolved, where the pointer is replaced with an explicit entity name (e.g., “author of 1984” → “George Orwell”); Paraphrased, where the reference is reworded but remains unresolved (e.g., “the writer of the dystopian novel 1984”); Removed, where the reference is deleted; and Kept Unchanged, where it is preserved verbatim.

\paragraph{\textbf{Results:}}
Operating on a fixed sample of 100 test queries (SEED = 42), we observe that the reformulation process overwhelmingly favors paraphrasing over explicit resolution. As shown in Table~\ref{tab:indirect_resolution}, for the Mistral model 94.68\% of indirect references are paraphrased, while only 1.06\% are resolved to a concrete entity name; 2.13\% are removed and 2.13\% remain unchanged. This yields a high unresolved ratio, indicating that the model generally refrains from committing to specific entity instantiations when indirect references are present.

\begin{table}[t]
\centering
\small
\caption{Distribution of reformulation behaviors for indirect entity references on a 100-query sample (SEED = 42) using Mistral.}
\label{tab:indirect_resolution}
\begin{tabular}{l c}
\toprule
\textbf{Reformulation Outcome} & \textbf{Percentage (\%)} \\
\midrule
Paraphrased     & 94.68\% \\
Resolved       & 1.06\% \\
Removed        & 2.13\% \\
Kept Unchanged & 2.13\% \\
\bottomrule
\end{tabular}
\end{table}

\paragraph{\textbf{Discussion:}}
This behavior is consistent with the design principles of our guardrailed entity expansion strategy. Unlike KB-based entity expansion approaches, where indirect references are aggressively resolved by linking to canonical identifiers in external resources such as Wikidata \cite{10.1145/2629489} or Google Knowledge Graph \cite{google_knowledge_graph_api}, our method explicitly restricts expansion to information that is unambiguously supported by the input query. As a result, when the prompt does not contain sufficient evidence to justify a specific entity instantiation, the model defaults to paraphrasing rather than speculative resolution.

Importantly, this conservatism should not be interpreted as a failure of reformulation, but rather as a deliberate trade-off aligned with open-world ToT retrieval. In many realistic scenarios, the intended target may be private, emerging, domain-specific, or absent from public knowledge bases altogether. In such cases, KB-dependent resolution would either fail silently or introduce hallucinated entities, tightly coupling system behavior to the coverage and biases of the underlying knowledge graph. By contrast, prompt-internal paraphrasing preserves semantic intent while avoiding overcommitment, allowing downstream retrieval to operate over a broader hypothesis space.

Taken together, these findings reinforce the distinction between cognitive reconstruction and knowledge-base completion. Our reformulation strategy prioritizes faithfulness and robustness over maximal entity resolution, ensuring that entity expansion remains query-grounded rather than knowledge-base driven. This design choice supports reliable behavior under ambiguity and distribution shift, which are defining characteristics of ToT retrieval settings.

\input{tables/rewriter-efficacy-dev3}

%% file: tables/rewriter-efficacy-dev3.tex
\begin{table*}[t]
\centering
\caption{Retrieval Performance Across Query Versions on the Pseudo-Test Set (dev3) using BM25.}
\label{tab:rewrite_efficacy}
\begin{tabularx}{\textwidth}{X c c c c c}
\toprule
\textbf{Query Version} &
\textbf{nDCG@10} &
\textbf{Recall@1000} &
\textbf{Success@100} &
\textbf{MRR} &
\textbf{MAP@10} \\
\midrule

Original & 0.3366 & 0.7705 & 0.6007 & 0.3129 & 0.3055 \\

\midrule

RM3 + BM25 & 0.2704 & 0.5093 & 0.3526 & 0.2663 & 0.2638 \\
Bo1 + BM25 & 0.3034 & 0.7295 & 0.5597 & 0.2798 & 0.2731 \\
KL + BM25  & 0.3093 & 0.7332 & 0.5690 & 0.2854 & 0.2785 \\

\midrule

Qwen   & 0.3720 & 0.7836 & 0.6474 & 0.3455 & 0.3382 \\
Llama  & 0.3889 & 0.7836 & 0.6549 & 0.3597 & 0.3533 \\
Mistral
& \textcolor{ForestGreen}{\textbf{0.4609}}
& \textcolor{ForestGreen}{\textbf{0.8284}}
& \textcolor{ForestGreen}{\textbf{0.7183}}
& \textcolor{ForestGreen}{\textbf{0.4338}}
& \textcolor{ForestGreen}{\textbf{0.4275}} \\

\bottomrule
\end{tabularx}
\end{table*}

%% file: results-analysis/bi-encoder-analysis.tex
\subsection{Analysis of Re-ranking by Bi-Encoder and Late-Interaction Models}

Following the sparse retrieval stage, we employed a Bi-encoder and Late Interaction Based Re-ranking ensemble consisting of ColBERTv2.0, Contriever, and E5-large. These models were combined using Reciprocal Rank Fusion (RRF) with a constant $k=60$.

To ensure computational efficiency without sacrificing coverage, we first performed the Robust Candidate Depth Tuning thoroughly discussed in \textbf{(Section \ref{robust-candidate-depth-tuning})}.

\subsubsection{\textbf{Optimal Candidate Depth Selection :}}

The depth tuning process revealed a decisive plateau in performance across all three bi-encoder models. Increasing the candidate depth ($K$) from 1,000 to 5,000 yielded zero improvement in the Robust Score or Mean Recall. This phenomenon confirms that the "recall ceiling" is strictly determined by the top-1000 documents retrieved by the sparse stage. Consequently, we selected \textbf{$K=1000$} as the optimal depth. Fetching candidates beyond this point incurs computational cost with zero potential gain in relevant document coverage.

\subsubsection{\textbf{The Sparse vs. Bi-Encoder Trade-off :}}

Upon executing the bi-encoder stage with $K=1000$ and fusing the results, we observed a specific phenomenon often described as the "Sparse Ceiling". \textbf{(Table \ref{tab:bi-encoder-degradation})} compares the performance of the input list (Mistral BM25) against the output list (Dense Fusion) on the pseudo-test set.

\input{tables/bi-encoder-degradation}

This data highlights a critical divergence between Ranking Quality and Recall Coverage:

\begin{enumerate}
\item \textbf{Ranking Degradation:} The Bi-Encoder and Late Interaction Re-ranking stage significantly degraded the ranking quality, dropping nDCG@10 by 18\%. This likely stems from the nature of the rewritten queries. Mistral optimized the ToT queries for lexical overlap (keywords), which BM25 exploits perfectly. The Dense models, trained on natural semantic matching, likely penalized these "keyword-stuffed" queries, down-ranking exact matches in favor of semantically broad but partially incorrect documents.
\end{enumerate}

\textbf{Strategic Decision:} While the drop in nDCG is suboptimal, the subsequent Cross-Encoder stage is explicitly designed to repair ranking errors within a high-recall pool. By proceeding with the RRF lists, we aim to combine the semantic signals captured by the bi-encoders and late-interaction mechanism with the precision of the Cross-Encoder, relying on the fact that the relevant documents are still present within the candidate pool.

%% file: tables/bi-encoder-degradation.tex
\begin{table}[t]
\centering
\caption{Performance comparison between the Input (Sparse) and Output (Dense Fusion) stages on the dev3 split.}
\label{tab:bi-encoder-degradation}
\begin{tabular}{c c c c}
\hline
\textbf{Metric} & \textbf{Mistral BM25} & \textbf{Bi-Encoder Fusion} & \textbf{Change} \\
\hline
nDCG@10    & \textcolor{ForestGreen}{\textbf{0.4609}} & 0.3778 & -18.0\% \\
Success@100 & 0.7183 & \textcolor{ForestGreen}{\textbf{0.7724}} & +7.5\% \\
Success@10 & \textcolor{ForestGreen}{\textbf{0.5672}} & 0.5448 & -3.9\% \\
Map@10 & \textcolor{ForestGreen}{\textbf{0.4275}} & 0.3249 & -24.0\% \\
MRR     & \textcolor{ForestGreen}{\textbf{0.4338}}   &  0.3352 & \textbf{-22.7\%} \\
\hline
\end{tabular}
\end{table}

%% file: results-analysis/cross-encoder-analysis.tex
\subsection{Analysis of Cross-Encoder Reranking}

To address the ranking degradation observed in the previous stage, we deployed a Cross-Encoder (monoT5) to re-rank the fused candidate lists. Unlike bi-encoders, which rely on compressed vector representations, the cross-encoder processes the query and document simultaneously, allowing for a fine-grained assessment of semantic relevance.

\subsubsection{\textbf{Performance Recovery and State-of-the-Art Results :}}

The results on the pseudo-test set (\texttt{dev3}) demonstrate a "Phoenix-like" recovery. As detailed in \textbf{(Table \ref{tab:ce_gain})}, the Cross-Encoder not only repaired the performance loss incurred during the dense stage but significantly outperformed the initial sparse baseline.

\input{tables/cross-encoder-gain}

Two critical insights emerge from this data:

\begin{itemize}
\item \textbf{Correction of Semantic Confusion:} The Bi-encoder and Late Interaction Based models struggled with the keyword-heavy nature of the rewritten queries, often burying relevant documents outside the top-10. The monoT5 model successfully identified these relevant pairs within the deeper candidate pool and elevated them to the top ranks, boosting Success@10 from 0.54 to 0.65.
\item \textbf{Recall Trade-off:} The total recall dropped from 0.828 to 0.733. This is an expected consequence of truncating the list to a fixed depth for re-ranking. However, the trade-off is highly favorable: we sacrifice the long-tail recall (documents ranked >300) to secure a massive improvement in top-tier ranking quality.
\end{itemize}

\subsubsection{\textbf{Optimization of Re-ranking Depth :}}

Given the high computational cost of Cross-Encoder inference, optimizing the candidate depth ($K$) is crucial for pipeline latency. We performed a robust tuning sweep to identify the smallest $K$ that maximizes ranking quality without introducing instability.

Using the same approach we used for the bi-encoder stage (Section \ref{robust-candidate-depth-tuning}), but instead tuning on nDCG@10, the robust score peaks at a remarkably shallow depth of \textbf{$K=50$}. This finding indicates that the "signal" for the Cross-Encoder is concentrated in the very top portion of the Dense Fusion list. Re-ranking deeper candidates (100--300) introduces more noise than relevance, diluting the robust score. Consequently, we configured the final pipeline to re-rank only the top 50 candidates, ensuring maximum efficiency for the final LLM generation stage.

%% file: tables/cross-encoder-gain.tex
\begin{table*}[t]
\centering
\caption{Performance comparison between the Dense Baseline (Input) and Cross-Encoder (Output) on the dev3 split.}
\label{tab:ce_gain}
\begin{tabular}{c c c c c c}
\hline
\textbf{Metric} 
& \textbf{Sparse Baseline}
& \textbf{Bi-Encoder Baseline} 
& \textbf{Cross-Encoder} 
& \textbf{Gain vs Bi-Encoder Baseline} 
& \textbf{Gain vs Sparse Baseline} \\
\hline
nDCG@10
& 0.4609
& 0.3778 
& \textcolor{ForestGreen}{\textbf{0.5196}} 
& +37.5\% 
& +12.7\% \\

Success@10
& 0.5672
& 0.5448 
& \textcolor{ForestGreen}{\textbf{0.6549}} 
& +20.2\% 
& +15.5\% \\

Map@10
& 0.4275
& 0.3249 
& \textcolor{ForestGreen}{\textbf{0.4765}} 
& +46.7\% 
& +11.5\% \\

MRR
& 0.4338
& 0.3352 
& \textcolor{ForestGreen}{\textbf{0.4808}}          
& +43.4\% 
& +10.8 \% \\
\hline
\end{tabular}
\end{table*}

%% file: results-analysis/llm-listwise-reranking-analysis.tex
\subsection{Analysis of List-wise LLM Re-ranking}

To achieve the final tier of ranking precision, we employed Qwen-2.5-72B Instruct AWQ as a List-wise Re-ranker. Unlike the point-wise or pair-wise scoring of previous stages, the LLM evaluates the candidate list holistically, leveraging its broad world knowledge to resolve the complex descriptions typical of ToT queries.

\subsubsection{\textbf{Performance on the Pseudo-Test Data :}}

The deployment of the LLM reranker resulted in a dramatic performance leap on the blind test set (\texttt{dev3}), establishing a new state-of-the-art benchmark for this pipeline. As shown in \textbf{(Table \ref{tab:llm_gain})}, the LLM successfully refined the Cross-Encoder's output, achieving an nDCG@10 of \textbf{0.6347}.

\input{tables/llm-reranking-gain}

The \textbf{29.2\% boost in MRR} is particularly telling. It indicates that Qwen-72B is exceptionally effective at identifying the ground truth within the top candidates and moving it to the Rank-1 position-a nuance that the smaller Cross-Encoder often missed.

\subsubsection{\textbf{Joint Parameter Tuning :}}

To optimize efficiency, we conducted a joint tuning sweep of the candidate window size passed from the Cross-Encoder ($K_{cross}$) and the depth of the LLM reranking ($K_{llm}$).

The results indicated that the performance saturates quickly. As observed in the tuning logs, smaller window sizes ($K_{cross}=30$) frequently outperformed or matched larger windows ($K_{cross}=50+$). This validates the efficiency of the pipeline: the relevant signals are concentrated at the very top. Forcing the LLM to process deeper ranks (e.g., rank 31-100) introduces noise rather than signal. Consequently, we configured the final stage with \textbf{$K_{cross}=30$} and \textbf{$K_{llm}=10$}, balancing maximum precision with computational cost.

%% file: tables/llm-reranking-gain.tex
\begin{table}[t]
\centering
\caption{Performance comparison between the Cross-Encoder (Input) and LLM Reranker (Output) on the pseudo-test set (dev3).}
\label{tab:llm_gain}
\begin{tabular}{c c c c}
\hline
\textbf{Metric} & \textbf{Cross-Encoder} & \textbf{LLM Reranker} & \textbf{Gain} \\
\hline
nDCG@10     & 0.5196 & \textcolor{ForestGreen}{\textbf{0.6347}} & +22.1\% \\
Success@10 & 0.6549 & \textcolor{ForestGreen}{\textbf{0.6828}} & +4.3\%  \\
Map@10 & 0.4765 & \textcolor{ForestGreen}{\textbf{0.6190}} & +29.9\%  \\
MRR        & 0.4808 & \textcolor{ForestGreen}{\textbf{0.6213}} & +29.2\% \\
\hline
\end{tabular}
\end{table}

%% file: final-pipeline-analysis/final-pipeline-analysis.tex
\section{Final Evaluation and Pipeline Analysis}

To validate the efficacy of our proposed architecture, we evaluated the complete multi-stage pipeline on the held-out test set. This split proved to be significantly more challenging than the development sets, characterized by lower absolute recall baselines and higher query ambiguity. Despite these challenges, the results confirm that a cascaded retrieval approach powered by the rewriting strategy is essential for solving Tip-of-the-Tongue (ToT) tasks.

\input{final-pipeline-analysis/overall-performance-trajectory}
\input{final-pipeline-analysis/stage-wise-analysis}

%% file: final-pipeline-analysis/overall-performance-trajectory.tex
\subsection{Overall Performance Trajectory}

The experimental results demonstrate a consistent, step-by-step improvement in ranking quality as the pipeline progresses from sparse retrieval to LLM powered list-wise re-ranking. As summarized in \textbf{(Table \ref{tab:final_pipeline_grouped})}, no single model is sufficient; rather, each stage contributes a specific layer of refinement.

\input{tables/final-pipeline-performance}

The final system achieves an nDCG@10 of \textbf{0.3706}, representing a \textbf{+113.21\% gain} over the sparse baseline. This dramatic lift validates our hypothesis: while lexical matching provides a necessary foundation, only deep semantic interaction (Cross-Encoders) and world-knowledge reasoning (LLMs) can bridge the vocabulary gap inherent in ToT queries.

\textbf{From Group 1 in Table~\ref{tab:final_pipeline_grouped}}, we observe that even a strong multi-stage retrieval and re-ranking pipeline performs poorly when applied directly to raw ToT queries. The primary failure mode occurs at the first retrieval stage, where Recall@1000 drops by 20.67\%. Specifically, the BM25+KL configuration achieves a maximum Recall@1000 of 0.5225 using the original queries, compared to 0.6302 when operating on Mistral-rewritten queries. This substantial loss of candidate coverage prevents downstream re-rankers from recovering relevant documents, since they are constrained to re-ranking only the top-1000 results produced by the initial retriever. Consequently, this early-stage recall deficit propagates through the pipeline, yielding overall gains of 33.88\% in nDCG@10, 29.92\% in MRR, and 29.98\% in MAP@10 when rewritten queries are used instead of the raw inputs.

%% file: tables/final-pipeline-performance.tex
\begin{table*}[t]
\centering
\caption{Cumulative performance of the retrieval pipeline on the test set, grouped by pipeline stage and query formulation.}
\label{tab:final_pipeline_grouped}

\setlength{\tabcolsep}{6pt}

\textbf{Table Group 1: Using Original Queries (Maximum Recall@1000 was recorded for the BM25 + KL combination (0.5225), which was used for downstream re-ranking }

\vspace{4pt}

\begin{tabularx}{\textwidth}{X c c c c c}
\toprule
\textbf{Model / Method} &
\textbf{nDCG@10} &
\textbf{MRR} &
\textbf{Success@10} &
\textbf{MAP@10} &
\textbf{Gain (nDCG@10)} \\
\midrule

\multicolumn{6}{l}{\textit{\textbf{Sparse Retrieval (Stage 1)}}} \\
\hspace*{6pt}BM25 & 0.1223 & 0.1114 & 0.3039 & 0.1064 & 0.00\% \\
\hspace*{6pt}BM25 + RM3 & 0.1033 & 0.0978 & 0.2363 & 0.0939 & --15.53\% \\
\hspace*{6pt}BM25 + Bo1 & 0.1136 & 0.1002 & 0.3103 & 0.0950 & --7.11\% \\
\hspace*{6pt}BM25 + KL & 0.1154 & 0.1030 & 0.3135 & 0.0974 & --5.64\% \\
\midrule

\multicolumn{6}{l}{\textit{\textbf{Bi-encoder Re-ranking (Stage 2)}}} \\
\hspace*{6pt}Contriever     & 0.1021 & 0.0896 & 0.3312 & 0.0832 & --11.52\% \\
\hspace*{6pt}ColBERTv2      & 0.0723 & 0.0654 & 0.2878 & 0.0588 & --37.34\% \\
\hspace*{6pt}E5-Large-v2    & 0.1837 & 0.1660 & 0.4228 & 0.1596 & +59.18\% \\
\hspace*{6pt}Dense Fusion (RRF) & 0.1579 & 0.1394 & 0.4068 & 0.1325 & +36.82\% \\
\midrule

\multicolumn{6}{l}{\textit{\textbf{Cross-Encoder Re-ranking (Stage 3)}}} \\
\hspace*{6pt}MonoT5 & 0.1839 & 0.1637 & 0.4164 & 0.1577 & +16.46\% \\
\midrule

\multicolumn{6}{l}{\textit{\textbf{List-wise LLM Re-ranking (Stage 4)}}} \\
\hspace*{6pt}Qwen 2.5 72B Instruct (4-bit) & 0.2768 & 0.2680 & 0.4180 & 0.2645 & +50.51\% \\

\midrule
\textbf{Net} & -- & -- & -- & -- & +139.86\% \\
\bottomrule
\end{tabularx}
\end{table*}

\begin{table*}[t]
\ContinuedFloat
\centering
\textbf{Table Group 2: Using Mistral Rewritten Queries (Ours). The first stage Recall@1000 is 0.6302, which already nets 20.61\% more relevant documents, than the best performing PRF using the original queries}

\vspace{4pt}
\setlength{\tabcolsep}{6pt}

\begin{tabularx}{\textwidth}{X c c c c c}
\toprule
\textbf{Model / Method} &
\textbf{nDCG@10} &
\textbf{MRR} &
\textbf{Success@10} &
\textbf{MAP@10} &
\textbf{Gain (nDCG@10)} \\
\midrule

\multicolumn{6}{l}{\textit{\textbf{Sparse Retrieval (Stage 1)}}} \\
\hspace*{6pt}BM25 
& 0.1738 & 0.1586 & 0.2476 & 0.1515 & 0.00\% \\
\midrule

\multicolumn{6}{l}{\textit{\textbf{Bi-encoder Re-ranking (Stage 2)}}} \\
\hspace*{6pt}Contriever     
& 0.1458 & 0.1283 & 0.2331 & 0.1188 & --16.15\% \\
\hspace*{6pt}ColBERTv2      
& 0.1361 & 0.1256 & 0.1961 & 0.1176 & --21.73\% \\
\hspace*{6pt}E5-Large-v2    
& 0.2253 & 0.1989 & 0.3376 & 0.1905 & +29.59\% \\
\hspace*{6pt}Dense Fusion (RRF) 
& 0.2338 & 0.2081 & 0.3457 & 0.1987 & +34.48\% \\
\midrule

\multicolumn{6}{l}{\textit{\textbf{Cross-Encoder Re-ranking (Stage 3)}}} \\
\hspace*{6pt}MonoT5 
& 0.3110 & 0.2822 & 0.4244 & 0.2757 & +33.04\% \\
\midrule

\multicolumn{6}{l}{\textit{\textbf{List-wise LLM Re-ranking (Stage 4)}}} \\
\hspace*{6pt}\textcolor{ForestGreen}{\textbf{Qwen 2.5 72B Instruct (4-bit)}} 
& \textcolor{ForestGreen}{\textbf{0.3706}}
& \textcolor{ForestGreen}{\textbf{0.3482}}
& \textcolor{ForestGreen}{\textbf{0.4550}}
& \textcolor{ForestGreen}{\textbf{0.3438}}
& \textcolor{ForestGreen}{\textbf{+19.16\%}} \\

\midrule
\textbf{Net} & -- & -- & -- & -- & \textbf{+113.21\%} \\
\textbf{Net Gain over Original} & -- & -- & -- & -- & \textbf{+33.88\%} \\

\bottomrule
\end{tabularx}

\end{table*}

%% file: final-pipeline-analysis/stage-wise-analysis.tex
\subsection{Stage-by-Stage Analysis}

\subsubsection{\textbf{Stage 1: The Sparse Foundation :}}
The BM25 baseline, augmented with Mistral-based query rewriting, achieved a Recall@1000 of 0.6302. Compared to the development sets (where recall often exceeded 0.80), this lower ceiling indicates that the test set contains "hard" queries with severe vocabulary mismatches that even generative rewriting could not fully resolve. This established a strict upper bound for retrieval coverage in subsequent stages.

\subsubsection{\textbf{Stage 2: Bi-encoder and Late Interaction Based Re-ranking and The "E5 Factor" :}}
The Bi-encoder and Late Interaction Based Re-ranking stage acted as a critical recovery mechanism, boosting nDCG@10 to 0.2338. However, the performance distribution among the constituent dense models was highly uneven:

\begin{itemize}
\item \textbf{ColBERTv2 (0.1361) and Contriever (0.1458):} These models underperformed the sparse baseline. This suggests a "semantic drift" where the models, confused by the abstract nature of the rewritten queries, retrieved documents that were topically related but factually incorrect.
\item \textbf{E5-Large (0.2253):} The E5 model significantly outperformed its peers, single-handedly driving the success of the dense stage. Its strong instruction-following capabilities likely aligned better with the structure of the rewritten queries.
\end{itemize}

The Reciprocal Rank Fusion (RRF) successfully combined these divergent signals, producing a final list (0.2338) that outperformed even the best single model (E5), justifying the computational overhead of the ensemble.

\subsubsection{\textbf{The Recall Ceiling Phenomenon :}} This is a structural consequence of our "Sparse-First Cascade" design. The dense models were configured to re-rank the top-1000 candidates provided by BM25 rather than retrieving from the full corpus. Consequently, the previous stage recall was mathematically capped by the sparse stage recall. While this limits total coverage, it massively reduces the search space for the dense vectors, thereby leading to the creation of a latency vs accuracy trade-off

\subsubsection{\textbf{Stages 3 and 4: Precision Enhancement :}}
The final two stages acted as "Precision Enhancers", tasked with cleaning the noisy candidate list.
\begin{itemize}
\item \textbf{Cross-Encoder:} The MonoT5 model delivered the single largest jump in the pipeline, increasing nDCG@10 from 0.2338 to 0.3110. It successfully identified relevant documents that had been buried deep in the fusion list and promoted them to the top 10.
\item \textbf{LLM Reranker:} The 72B parameter model provided the final polish, achieving a further +19\% gain over the Cross-Encoder. By leveraging its extensive world knowledge, it rescued hard cases, pushing Success@10 to 0.4550, demonstrating an ability to solve queries that lacked sufficient semantic overlap for the smaller models to detect.
\end{itemize}

%% file: conclusion.tex
\section{Conclusion}

This work reframes Tip-of-the-Tongue (ToT) retrieval as a problem of cognitive signal reconstruction rather than conventional document ranking. Through extensive evaluation on the TREC-ToT 2025 benchmark, we demonstrate that even a strong multi-stage retrieval pipeline struggles when operating directly on raw ToT queries, confirming that query formulation constitutes the dominant bottleneck in this setting.

By introducing a lightweight, single-pass LLM-driven query reformulation and controlled expansion step prior to retrieval, we achieve substantial gains in both first-stage recall and downstream ranking effectiveness. Notably, this intervention requires no task-specific fine-tuning, no corpus-aware supervision, and no domain adaptation. All models are used off-the-shelf in a zero-shot configuration, indicating that the observed improvements stem from prompt design and query reconstruction strategy rather than specialization of the underlying LLMs.

Our results show that modest pre-retrieval cognitive reconstruction unlocks large performance improvements across the entire pipeline, yielding consistent gains in Recall, nDCG@10, MRR, and MAP@10 relative to identical systems operating on original queries. These findings underscore the importance of treating query interpretation as a first-class component of retrieval, particularly in scenarios characterized by vague, partial, or inaccurate user memories.

More broadly, the proposed approach supports an open-world retrieval paradigm in which collections evolve and target domains are not known in advance. By decoupling effectiveness from corpus-specific training and emphasizing query-centric reasoning, the framework offers a practical and cost-effective path toward robust ToT retrieval in realistic deployment settings. We hope this work encourages further investigation of cognitive reconstruction as a foundational layer in future information retrieval systems.
\\\\
\textbf{Limitations and Future Work:} Several limitations point to promising directions for future research. First, our approach performs query reconstruction in a single pass; iterative refinement that incorporates retrieval feedback may yield further gains, albeit at increased computational cost. Second, while we focus exclusively on text-based ToT queries, the effectiveness of cognitive reconstruction for multimodal ToT scenarios such as visual memories combined with partial textual descriptions remains unexplored. Finally, although we observe consistent improvements across three families of 8B-parameter LLMs, evaluating larger-capacity models (e.g., 72B or 120B) and extending experiments to additional languages would further strengthen claims of generalizability, particularly given the potential benefits of increased internal reasoning capacity. Finally, while our current system enforces a single, query-grounded expansion policy, an important direction for future work is adaptive entity expansion. In practice, different deployment settings impose different constraints: fixed-domain environments may benefit from explicit linking to external resources such as Wikidata \cite{10.1145/2629489} or Google Knowledge Graph \cite{google_knowledge_graph_api}, whereas open-world or private corpora require strict grounding to the input query alone. A promising extension of our framework would be to automatically select between KB-backed resolution and prompt-internal, query-faithful paraphrasing based on the target domain and data regime, enabling the same cognitive reconstruction pipeline to operate robustly across closed-world, open-world, and proprietary settings.